\documentclass[aps,pre,draft,showpacs,10pt]{revtex4}
\usepackage{amsmath}

\begin{document}

\title{First Principles Derivation of Effective Ginzburg-Landau Free Energy Models for Crystalline Systems}
\author{James F. Lutsko}
\affiliation{Center for Nonlinear Phenomena and Complex Systems, Universit\'{e} Libre de
Bruxelles, C.P. 231, Blvd. du Triomphe, 1050 Brussels, Belgium}
\date{\today }

\begin{abstract}
The expression of the free energy density of a classical
crystalline system as a gradient expansion in terms of a set of order
parameters is developed using classical density functional theory. The goal here is to extend and complete an earlier
derivation by L{\"o}wen et al (Europhys. Lett.9, 791, 1989). The
limitations of the resulting expressions are also discussed including the
boundary conditions needed for finite systems and the fact that the results
cannot, at present, be used to take into account elastic relaxation.
\end{abstract}

\maketitle

\section{Introduction}

Classical Density Functional Theory (DFT)\ can be used to determine the
thermodynamic and structural properties of classical systems based on only
the interaction potential of the atoms and the external fields (including
those of the walls if any) they are subject to\cite{EvansDFT}. It is based
on the fact that the Helmholtz free energy is a unique functional of the
local density and the goal of practical formulations of DFT is to provide an
approximate form for that functional relation that can be used to calculate
the properties of real systems under a variety of thermodynamic and
geometric conditions\cite{EvansDFT},\cite{HansenMcdonald}. For inhomogeneous
systems, containing e.g. surfaces or bulk interfaces, even the simplest
approximate functionals can be analytically complex to work with so that
from the earliest days of the development of DFT in the 1970's it has been
common in such circumstances to use gradient expansions in the order
parameters in order to obtain simplified forms of the theory\cite{EvansDFT}.
In fact, two types of quite different gradient expansions are currently in
use. The first, described by Evans\cite{EvansDFT} and refined by Haymet and
Oxtoby\cite{OxtobyHaymet1},\cite{OxtobyHaymet2}, yields equations for the
local density which vary with the microscopic structure of the system: for
example, when applied to a solid, that some terms in the equations will
exhibit variations on the length scale of the lattice parameter. The second
type of gradient expansion is that of L\"{o}%
wen, Beier and Wagner\cite{Lowen1,Lowen2} which yields a typical Ginzburg-Landau model free energy. While based on similar ideas to
the first method, this results in a true long-wavelength approximation in
which microscopic details, such as the underlying lattice when applied to a
solid, are subsumed into the various coefficients of the gradient expansion.
The resulting simplified free energy functional therefore takes the form of
a typical phenomenological Landau theory, but with coefficients that can be
calculated from microscopic information. This theory has been used to
discuss surface melting, the nucleation of freezing in a bulk fluid\cite{OLW}%
,\cite{vwdCrystalMelt} and solid-solid phase transitions\cite{vdwSolidSolid}.

The purpose of this paper is twofold. The first goal is to review the two
types of gradient expansions in order to clarify the differences between
them. The need to do so is motivated by certain criticisms that have been
made in the literature\cite{DftGradientCriticism} about the inadequacy of
the gradient expansions:\ as shown below, these seem likely to be relevant
to the older class but not to the Ginzburg-Landau model. The second goal
is to clarify, extend and complete the derivation of the Ginzburg-Landau models. The derivation currently in the literature\cite{Lowen1},\cite%
{Lowen2} was developed only for planar interfaces and is also incomplete.
Furthermore, issues such as the appropriate boundary conditions for finite
systems have not been addressed.

In the next Section, the required elements of classical DFT are reviewed and
the older form of the gradient expansion is discussed. Section III presents
in detail a derivation of the Ginzburg-Landau model of L{\"o}wen et al.
Finally, in the last Section the differences between the two types of
expansion are discussed, as are the advantages and limitations of the models.

\section{Density Functional Theory}

Consider a system of $N$ atoms with positions and momenta $%
\overrightarrow{q}_{i}$ and $\overrightarrow{p}_{i}$ respectively and mass $%
m_{i}$. The atoms interact via a pair potential $v\left( \overrightarrow{r}%
\right) $ and are also subject to the effect of an external field $\phi
\left( \overrightarrow{r}\right) $ which includes any one-body forces such
as gravity or applied electrical fields as well as the effects of any walls.
The local number density is 
\begin{equation}
\widehat{\rho }\left( \overrightarrow{r}\right) =\sum_{i=1}^{N}\delta \left( 
\overrightarrow{r}-\overrightarrow{q}_{i}\right)   \label{dft1}
\end{equation}%
and its average in a grand-canonical ensemble is denoted%
\begin{equation}
\rho \left( \overrightarrow{r}\right) =\left\langle \widehat{\rho }\left( 
\overrightarrow{r}\right) \right\rangle .  \label{dft2}
\end{equation}%
The grand distribution is 
\begin{equation}
f=\Xi ^{-1}\frac{1}{N!h^{3N}}\exp \left( -\beta H+\beta \mu N\right) 
\label{dft3}
\end{equation}%
where $h$ is Planck's constant, $\beta =1/k_{B}T$ is the inverse
temperature, $\mu $ is the chemical potential and the Hamiltonian is 
\begin{equation}
H=\sum_{i=1}^{N}\frac{1}{2m_{i}}p_{i}^{2}+\sum_{i<j}v\left( \overrightarrow{q%
}_{ij}\right) +\int d\overrightarrow{r}\;\widehat{\rho }\left( 
\overrightarrow{r}\right) \phi \left( \overrightarrow{r}\right) .
\label{dft4}
\end{equation}%
The grand partition function is 
\begin{equation}
\Xi =\sum_{N}\int d\overrightarrow{q}^{N}d\overrightarrow{p}^{N}\;\frac{1}{%
N!h^{3N}}\exp \left( -\beta H+\mu N\right)   \label{dft5}
\end{equation}%
and it is related to the grand potential, $\Omega $, as 
\begin{equation}
\Omega =-k_{B}T\ln \Xi .  \label{dft6}
\end{equation}%
The key results of DFT\cite{EvansDFT},\cite{HansenMcdonald} are that (a) there is a one-to-one correspondence
between fields $\phi \left( \overrightarrow{r}\right) $ and average density
distributions $\rho \left( \overrightarrow{r}\right) $ leading to the fact
that (b) there exists a unique functional of the density $F\left[ \rho %
\right] $ with the property that the functional 
\begin{equation}
\Omega \left( \left[ n\right] ,\left[ \phi \right] \right) \equiv F\left[ n%
\right] -\mu \int d\overrightarrow{r}\;n\left( \overrightarrow{r}\right)
+\int d\overrightarrow{r}\;n\left( \overrightarrow{r}\right) \phi \left( 
\overrightarrow{r}\right)   \label{dft7}
\end{equation}%
is minimized, for a fixed field, by the equilibrium density%
\begin{equation}
\left. \frac{\delta \Omega }{\delta n}\right| _{\phi ,\mu ,T}\left( n=\rho
\right) =0  \label{dft8}
\end{equation}%
where the notation indicates that the variation is performed at fixed
external field $\phi $ as well as fixed chemical potential and temperature%
\cite{EvansDFT},\cite{HansenMcdonald}. The functional $F\left[ n\right] $ is
commonly written as $F\left[ n\right] =F_{id}\left[ n\right] +F_{ex}\left[ n%
\right] $ with the ideal contribution given by 
\begin{equation}
\beta F_{id}\left[ n\right] =\int d\overrightarrow{r}\;\left( n\left( 
\overrightarrow{r}\right) \ln \left( \Lambda ^{3}n\left( \overrightarrow{r}%
\right) \right) -n\left( \overrightarrow{r}\right) \right) ,  \label{dft9}
\end{equation}%
and where the excess contribution is in general unknown. However, it is
related to the $m$-body direct correlation functions by%
\begin{equation}
c_{n}\left( \overrightarrow{r}_{1},...,\overrightarrow{r}_{m}\right) =-\frac{%
\delta ^{m}\beta F_{ex}\left[ n\right] }{\delta n\left( \overrightarrow{r}%
_{1}\right) ...\delta n\left( \overrightarrow{r}_{m}\right) }.  \label{dft10}
\end{equation}%
Combining eqs.(\ref{dft7})-(\ref{dft10}), the Euler-Lagrange equation can be
written as%
\begin{equation}
\ln \left( \Lambda ^{3}n\left( \overrightarrow{r}\right) \right)
-c_{1}\left( \overrightarrow{r}\right) -\beta \mu +\beta \phi \left( 
\overrightarrow{r}\right) =0.  \label{dft12}
\end{equation}%
Finally, in terms of these quantities, the Helmholtz free energy $A$ is 
\begin{equation}
A=F\left[ \rho \right] +\int d\overrightarrow{r}\;\rho \left( 
\overrightarrow{r}\right) \phi \left( \overrightarrow{r}\right) .
\label{dft11}
\end{equation}

All practical uses of DFT are based on approximate functionals $F_{ex}\left[
n\right] $ motivated by various physical reasons. Given such an
approximation, one then has two choices as to how to proceed. The first is
to use the functional to derive the one-body direct correlation function
from eq.(\ref{dft10}) and then to derive the density from eq.(\ref{dft12}).
(Note that in bulk systems, the only external force of interest is often
that associated with the walls which contain the system. As this volume goes
to infinity, it is expected that interior points are unaffected by the
wall-force so that in these so-called ''self-sustaining'' systems, one can
take $\beta \phi \left( \overrightarrow{r}\right) =0$.)\ The second approach
is to parameterize the density and to then determine the parameters via eq(%
\ref{dft8}). To be specific, let the parameterized density be $n\left( 
\overrightarrow{r}\right) =\rho \left( \overrightarrow{r};\Gamma \right) $
where $\Gamma =\left\{ \Gamma _{a}\right\} $ represents a family of
parameters. In a bulk system, the parameters are constants and eq.(\ref{dft8}%
) implies that%
\begin{equation}
\frac{\partial F}{\partial \Gamma _{a}}-\mu \frac{\partial N}{\partial
\Gamma _{a}}+\int d\overrightarrow{r}\;\frac{\partial \rho \left( 
\overrightarrow{r};\Gamma \right) }{\partial \Gamma _{a}}\phi \left( 
\overrightarrow{r}\right) =0.  \label{dft13}
\end{equation}%
If the parameters depended on position, the result would be%
\begin{equation}
\left[ \frac{\delta F}{\delta n\left( \overrightarrow{r}\right) }\right]
_{\rho \left( \overrightarrow{r};\Gamma \right) }\frac{\partial \rho \left( 
\overrightarrow{r};\Gamma \right) }{\partial \Gamma _{a}\left( 
\overrightarrow{r}\right) }-\mu \frac{\partial \rho \left( \overrightarrow{r}%
;\Gamma \right) }{\partial \Gamma _{a}\left( \overrightarrow{r}\right) }%
+\phi \left( \overrightarrow{r}\right) \frac{\partial \rho \left( 
\overrightarrow{r};\Gamma \right) }{\partial \Gamma _{a}\left( 
\overrightarrow{r}\right) }=0.  \label{dft14}
\end{equation}%
One trivial example of such a parameterization is a bulk liquid in which
case one takes $\rho \left( \overrightarrow{r};\Gamma \right) =\rho _{0}$
for some constant $\rho _{0}$. Then, since the field is taken to be zero in
the interior, the Euler-Lagrange equations simply give the usual
thermodynamic relation%
\begin{equation}
\frac{\partial F}{\partial \rho _{0}}-\mu V=0.  \label{dft15}
\end{equation}%
A less trivial example is that of a bulk crystalline solid which is
typically parameterized by a sum of Gaussians giving%
\begin{equation}
\rho \left( \overrightarrow{r};\Gamma \right) =\sum_{i=1}^{N_{V}}\left( 
\frac{\alpha }{\pi }\right) ^{3/2}\exp \left( -\alpha \left( \overrightarrow{%
r}-\overrightarrow{R}_{i}\right) ^{2}\right)  \label{dft16}
\end{equation}%
where $\left\{ R_{i}\left( \rho _{0}\right) \right\} _{i=1}^{N_{v}}$ are the
Bravais lattice vectors within the volume $V$ and are, as indicated,
functions of the average density $\rho _{0}$. Then, the parameters are $%
\Gamma =\left\{ \alpha ,\rho _{0}\right\} $ and since the field is taken to
be zero in the interior, the Euler-Lagrange equations reduce to%
\begin{eqnarray}
\frac{\partial F}{\partial \alpha } &=&0  \label{dft17} \\
\frac{\partial F}{\partial \rho _{0}}-\mu V &=&0.  \notag
\end{eqnarray}%
A solid-liquid interfacial system could be described by taking, e.g., $%
\alpha $ to be spatially dependent and by using eq.(\ref{dft14}). Note that
this requires that the free energy functional $F\left[ n\right] $ be known
not only for bulk systems but for inhomogeneous systems as well. Given this
functional, the choice of whether to use a parameterized density or to
directly solve eq.(\ref{dft12}) is a matter of convenience.

\subsection{Gradient Expansion}

The classical gradient expansion of DFT was described by Evans\cite{EvansDFT}
and given a more modern derivation by Haymet and Oxtoby\cite{OxtobyHaymet1},%
\cite{OxtobyHaymet2}. Here, the latter derivation is reviewed so as to contrast
with the Ginzburg-Landau theory discussed below.

Equation(\ref{dft10}) can be functionally integrated in density space.
Define a parameterized $n_{\lambda }\left( \overrightarrow{r}\right) =%
\overline{n}_{0}+\lambda \left( n_{1}\left( \overrightarrow{r}\right) -%
\overline{n}_{0}\right) $where $n_{1}\left( \overrightarrow{r}\right) =\rho
\left( \overrightarrow{r};\Gamma \left( \overrightarrow{r}\right) \right) $
is the state of interest and $\overline{n}_{0}$ corresponds to any
convenient uniform (i.e. liquid) state.Then two integrations and adding and
subtracting the same expression for $n_{1}\left( \overrightarrow{r}\right) =%
\overline{n}_{1}$ gives%
\begin{eqnarray}
\beta F_{ex}\left[ n_{1}\right] -\beta F_{ex}\left( \overline{n}_{1}\right)
&=&-\int d\overrightarrow{r}_{1}d\overrightarrow{r}_{2}\int_{0}^{1}%
\int_{0}^{\lambda }c_{2}\left( \overrightarrow{r}_{1},\overrightarrow{r}%
_{2};n_{\lambda ^{\prime }}\right) \left( n_{1}\left( \overrightarrow{r}%
_{1}\right) -\overline{n}_{0}\right) \left( n_{1}\left( \overrightarrow{r}%
_{2}\right) -\overline{n}_{0}\right) d\lambda d\lambda ^{\prime }.  \notag \\
&&+\int d\overrightarrow{r}_{1}d\overrightarrow{r}_{2}\int_{0}^{1}\int_{0}^{%
\lambda }c_{2}\left( \overrightarrow{r}_{1},\overrightarrow{r}_{2};\overline{%
n}_{\lambda ^{\prime }}\right) \left( \overline{n}_{1}-\overline{n}%
_{0}\right) \left( \overline{n}_{1}-\overline{n}_{0}\right) d\lambda
d\lambda ^{\prime }.
\end{eqnarray}%
Adding in the ideal contribution gives%
\begin{eqnarray}
\beta F\left[ n_{1}\right] -\beta F\left( \overline{n}_{1}\right) &=&\int d%
\overrightarrow{r}_{1}\;\left( n_{1}\left( \overrightarrow{r}_{1}\right) \ln
n_{1}\left( \overrightarrow{r}_{1}\right) -\overline{n}_{1}\ln \overline{n}%
_{1}\right) \\
&&-\int d\overrightarrow{r}_{1}d\overrightarrow{r}_{2}\int_{0}^{1}\int_{0}^{%
\lambda }c_{2}\left( \overrightarrow{r}_{1},\overrightarrow{r}%
_{2};n_{\lambda ^{\prime }}\right) \left( n_{1}\left( \overrightarrow{r}%
_{1}\right) -\overline{n}_{0}\right) \left( n_{1}\left( \overrightarrow{r}%
_{2}\right) -\overline{n}_{0}\right) d\lambda d\lambda ^{\prime }.  \notag \\
&&+\int d\overrightarrow{r}_{1}d\overrightarrow{r}_{2}\int_{0}^{1}\int_{0}^{%
\lambda }c_{2}\left( \overrightarrow{r}_{1},\overrightarrow{r}_{2};\overline{%
n}_{\lambda ^{\prime }}\right) \left( \overline{n}_{1}-\overline{n}%
_{0}\right) \left( \overline{n}_{1}-\overline{n}_{0}\right) d\lambda
d\lambda ^{\prime },  \notag
\end{eqnarray}%
and it should be  noted that all of these relations are exact.

Haymet and Oxtoby then introduce the following approximations. First, the
two-body dcf is expanded about that of a uniform system as $c_{2}\left( 
\overrightarrow{r}_{1},\overrightarrow{r}_{2};n_{\lambda ^{\prime }}\right)
=c_{2}\left( r_{12};n_{0}\right) +...$and higher order terms dropped giving%
\begin{eqnarray}
\beta F\left[ n\right] -\beta F\left( \overline{n}_{1}\right) &\simeq &\int d%
\overrightarrow{r}_{1}\;\left( n_{1}\left( \overrightarrow{r}_{1}\right) \ln
n_{1}\left( \overrightarrow{r}_{1}\right) -\overline{n}_{1}\ln \overline{n}%
_{1}\right) \\
&&-\frac{1}{2}\int c_{2}\left( r_{12};n_{0}\right) \left( n_{1}\left( 
\overrightarrow{r}_{1}\right) -\overline{n}_{1}\right) \left( n_{1}\left( 
\overrightarrow{r}_{2}\right) +\overline{n}_{1}\right) d\overrightarrow{r}%
_{1}d\overrightarrow{r}_{2}  \notag
\end{eqnarray}%
Introducing the local free energy difference for a uniform system%
\begin{eqnarray}
\Delta \beta f\left( \overrightarrow{r}_{1};\Gamma _{0}\right) &=&\rho
_{1}\left( \overrightarrow{r}_{1};\Gamma _{0}\right) \ln \rho _{1}\left( 
\overrightarrow{r}_{1};\Gamma _{0}\right) -\overline{\rho }_{1}\ln \overline{%
\rho }_{1} \\
&&-\frac{1}{2}\int c_{2}\left( r_{12};n_{0}\right) \left( \rho _{1}\left( 
\overrightarrow{r}_{1};\Gamma _{0}\right) +\overline{\rho }_{1}\right)
\left( \rho _{1}\left( \overrightarrow{r}_{2};\Gamma _{0}\right) -\overline{%
\rho }_{1}\right) d\overrightarrow{r}_{2}  \notag
\end{eqnarray}%
the last result can be written more suggestively as%
\begin{eqnarray}
\beta F\left[ n\right] -\beta F\left( \overline{n}_{1}\right) &\simeq &\int d%
\overrightarrow{r}_{1}\;\Delta \beta f\left( \overrightarrow{r}_{1};\Gamma
\left( \overrightarrow{r}_{1}\right) \right)  \notag \\
&&-\frac{1}{2}\int c_{2}\left( r_{12};n_{0}\right) \left( \rho _{1}\left( 
\overrightarrow{r}_{1};\Gamma _{0}\right) +\overline{\rho }_{1}\right)
\left( \rho _{1}\left( \overrightarrow{r}_{2};\Gamma _{0}\right) -\overline{%
\rho }_{1}\right) d\overrightarrow{r}_{1}d\overrightarrow{r}_{2}
\end{eqnarray}%
so that the first term on the right has the form of a mean-field
approximation. The second term can be expanded to give, up to second order
in gradients of the order parameter%
\begin{eqnarray}
\beta F\left[ n\right] -\beta F\left( \overline{n}_{1}\right) &\simeq &\int d%
\overrightarrow{r}_{1}\;\Delta \beta f\left( \overrightarrow{r}_{1};\Gamma
\left( \overrightarrow{r}_{1}\right) \right) \\
&&-\frac{1}{2}\int c_{2}\left( r_{12};n_{0}\right) \left( \rho _{1}\left( 
\overrightarrow{r}_{1};\Gamma \left( \overrightarrow{r}_{1}\right) \right) +%
\overline{\rho }_{1}\right) \frac{\partial \rho _{1}\left( \overrightarrow{r}%
_{2};\Gamma \left( \overrightarrow{r}_{1}\right) \right) }{\partial \Gamma
\left( \overrightarrow{r}_{1}\right) }\overrightarrow{r}_{21}\cdot 
\overrightarrow{\nabla }\Gamma \left( \overrightarrow{r}_{1}\right) d%
\overrightarrow{r}_{1}d\overrightarrow{r}_{2}  \notag \\
&&-\frac{1}{4}\int c_{2}\left( r_{12};n_{0}\right) \left( \rho _{1}\left( 
\overrightarrow{r}_{1};\Gamma \left( \overrightarrow{r}_{1}\right) \right) +%
\overline{\rho }_{1}\right) \frac{\partial \rho _{1}\left( \overrightarrow{r}%
_{2};\Gamma \left( \overrightarrow{r}_{1}\right) \right) }{\partial \Gamma
\left( \overrightarrow{r}_{1}\right) }\overrightarrow{r}_{21}\overrightarrow{%
r}_{21}\cdot \overrightarrow{\nabla }\overrightarrow{\nabla }\Gamma \left( 
\overrightarrow{r}_{1}\right) d\overrightarrow{r}_{1}d\overrightarrow{r}_{2}
\notag \\
&&-\frac{1}{4}\int c_{2}\left( r_{12};n_{0}\right) \left( \rho _{1}\left( 
\overrightarrow{r}_{1};\Gamma \left( \overrightarrow{r}_{1}\right) \right) +%
\overline{\rho }_{1}\right) \frac{\partial _{1}^{2}\rho \left( 
\overrightarrow{r}_{2};\Gamma \left( \overrightarrow{r}_{1}\right) \right) }{%
\partial \Gamma ^{2}\left( \overrightarrow{r}_{1}\right) }\left( 
\overrightarrow{r}_{21}\cdot \overrightarrow{\nabla }\Gamma \left( 
\overrightarrow{r}_{1}\right) \right) ^{2}d\overrightarrow{r}_{1}d%
\overrightarrow{r}_{2}  \notag
\end{eqnarray}%
OH use a specific parameterization%
\begin{equation}
\rho _{1}\left( \overrightarrow{r};\Gamma \left( \overrightarrow{r}\right)
\right) =\rho _{0}\left( 1+\Gamma _{0}\left( \overrightarrow{r}\right)
+\sum_{n>0}\exp \left( i\overrightarrow{K}_{n}\cdot \overrightarrow{r}%
\right) \Gamma _{n}\left( \overrightarrow{r}\right) \right)  \label{p1}
\end{equation}%
where $\overrightarrow{K}_{n}$ is the $nth$ reciprocal lattice vector.
Substituting into the expression for the free energy gives%
\begin{eqnarray}
\beta F\left[ n_{1}\right] -\beta F\left( \overline{n}_{1}\right) &\simeq
&\int d\overrightarrow{r}_{1}\;\Delta \beta f\left( \overrightarrow{r}%
_{1};\Gamma \left( \overrightarrow{r}_{1}\right) \right) \\
&&-\frac{1}{2}\rho _{0}\int c_{2}\left( r_{12};\rho _{0}\right) \left( \rho
_{0}+n_{0}\Gamma _{0}\left( \overrightarrow{r}_{1}\right) +\overline{n}%
_{1}\right) \overrightarrow{r}_{21}\cdot \overrightarrow{\nabla }\Gamma
_{0}\left( \overrightarrow{r}_{1}\right) d\overrightarrow{r}_{1}d%
\overrightarrow{r}_{2}  \notag \\
&&-\frac{1}{2}\rho _{0}\sum_{n>0}\int c_{2}\left( r_{12};\rho _{0}\right)
\exp \left( i\overrightarrow{K}_{n}\cdot \overrightarrow{r}_{12}\right)
\Gamma _{n}\left( \overrightarrow{r}_{1}\right) \overrightarrow{r}_{21}\cdot 
\overrightarrow{\nabla }\Gamma _{n}\left( \overrightarrow{r}_{1}\right) d%
\overrightarrow{r}_{1}d\overrightarrow{r}_{2}  \notag \\
&&-\frac{1}{4}\rho _{0}\int c_{2}\left( r_{12};\rho _{0}\right) \left( \rho
_{0}+\rho _{0}\Gamma _{0}\left( \overrightarrow{r}_{1}\right) +\overline{%
\rho }_{1}\right) \overrightarrow{r}_{21}\overrightarrow{r}_{21}\cdot 
\overrightarrow{\nabla }\overrightarrow{\nabla }\Gamma _{0}\left( 
\overrightarrow{r}_{1}\right) d\overrightarrow{r}_{1}d\overrightarrow{r}_{2}
\notag \\
&&-\frac{1}{4}\rho _{0}\sum_{n>0}\int c_{2}\left( r_{12};\rho _{0}\right)
\exp \left( i\overrightarrow{K}_{n}\cdot \overrightarrow{r}_{12}\right)
\Gamma _{n}\left( \overrightarrow{r}_{1}\right) \overrightarrow{r}_{21}%
\overrightarrow{r}_{21}\cdot \overrightarrow{\nabla }\overrightarrow{\nabla }%
\Gamma _{n}\left( \overrightarrow{r}_{1}\right) d\overrightarrow{r}_{1}d%
\overrightarrow{r}_{2}.  \notag
\end{eqnarray}%
The linear terms vanish by symmetry of the uniform fluid and, after changing
variables and integrating by parts, the final result is%
\begin{eqnarray}
\beta F\left[ n_{1}\right] -\beta F\left( \overline{n}_{1}\right) &\simeq
&\int d\overrightarrow{r}_{1}\;\Delta \beta f\left( \overrightarrow{r}%
_{1};\Gamma \left( \overrightarrow{r}_{1}\right) \right) \\
&&+\frac{1}{4}\rho _{0}^{2}\int c_{2}\left( r_{12};\rho _{0}\right) \left( 
\overrightarrow{r}_{21}\cdot \Gamma _{0}\left( \overrightarrow{r}_{1}\right)
\right) ^{2}d\overrightarrow{r}_{1}d\overrightarrow{r}_{12}  \notag \\
&&+\frac{1}{4}\rho _{0}^{2}\sum_{n>0}\int c_{2}\left( r_{12};\rho
_{0}\right) \exp \left( i\overrightarrow{K}_{n}\cdot \overrightarrow{r}%
_{12}\right) \left( \overrightarrow{r}_{21}\cdot \overrightarrow{\nabla }%
\Gamma _{n}\left( \overrightarrow{r}_{1}\right) \right) ^{2}d\overrightarrow{%
r}_{1}d\overrightarrow{r}_{12}.  \notag
\end{eqnarray}%
This can be written in the form%
\begin{equation}
\beta F\left[ n_{1}\right] -\beta F\left( \overline{n}_{1}\right) \simeq
\int d\overrightarrow{r}\;\left[ \Delta \beta f\left( \overrightarrow{r}%
;\Gamma \left( \overrightarrow{r}\right) \right) +\frac{1}{2}%
\sum_{ab}K_{ij}^{ab}\left( \partial _{i}\Gamma _{a}\left( \overrightarrow{r}%
\right) \right) \left( \partial _{j}\Gamma _{b}\left( \overrightarrow{r}%
\right) \right) \right]  \label{EOH}
\end{equation}%
with%
\begin{equation}
K_{ij}^{ab}=\delta _{ab}\frac{1}{2}\rho _{0}^{2}\int \exp \left( i%
\overrightarrow{K}_{a}\cdot \overrightarrow{r}\right) r_{i}r_{j}c_{2}\left(
r;n_{0}\right) d\overrightarrow{r}.
\end{equation}

The main criticisms of this theory are that (a) all correlation functions
are eventually expressed in terms of an expansion about the uniform liquid
and these expansions truncated at the two-body dcf; (b) the latter stages of
the derivation depend on a particular parameterization of the density and
(c) the mean field contribution, $\Delta f\left( \overrightarrow{r}%
_{1};\Gamma \left( \overrightarrow{r}_{1}\right) \right) $, is not a truly
long-wavelength approximation as it clearly exhibits variations on the
length scale of the lattice. To clarify the last point: in a long-wavelength
theory, one would expect to see the term%
\begin{equation}
\int d\overrightarrow{r}_{1}\;\Delta \beta f\left( \overrightarrow{r}%
_{1};\Gamma \left( \overrightarrow{r}_{1}\right) \right)
\end{equation}%
replaced by something like%
\begin{equation}
\int d\overrightarrow{R}\int d\overrightarrow{r}_{1}\;\Delta \beta f\left( 
\overrightarrow{r}_{1};\Gamma \left( \overrightarrow{R}\right) \right) =\int
d\overrightarrow{R}\;\Delta \beta F\left( \Gamma \left( \overrightarrow{R}%
\right) \right) \;
\end{equation}%
so that the variational equations for the order parameters do not contain
terms that vary over microscopic length scales.

\section{Effective Ginzburg-Landau model}

Ideally, one would like to obtain a Helmholtz free energy in which
microscopic length scales do not occur of the generic form%
\begin{equation}
\beta F\left[ \rho \right] =\int d\overrightarrow{r}\;\left[ f_{0}\left(
\Gamma \left( \overrightarrow{r}\right) \right) +\frac{1}{2}%
K_{ij}^{ab}\left( \Gamma \left( \overrightarrow{r}\right) \right) \left( 
\frac{\partial }{\partial r_{i}}\Gamma _{a}\left( \overrightarrow{r}\right)
\right) \left( \frac{\partial }{\partial r_{j}}\Gamma _{b}\left( 
\overrightarrow{r}\right) \right) +...\right]  \label{Target}
\end{equation}%
in which dependence on the atomic structure has been ''integrated out''.
Furthermore, one would expect the first term to take the typical mean-field
form of the bulk free energy evaluated for the local order parameters,%
\begin{equation}
Vf_{0}\left( \Gamma \left( \overrightarrow{r}_{1}\right) \right) =\left.
\beta F\left[ \rho \right] \right| _{\Gamma =\Gamma \left( \overrightarrow{r}%
_{1}\right) }\equiv \beta F_{bulk}\left( \Gamma \left( \overrightarrow{r}%
\right) \right) ,
\end{equation}%
where $\beta \Phi _{bulk}\left( \Gamma _{0}\right) $ is the free energy of a
bulk system with spatially uniform order parameters $\Gamma _{0}$. Given a
free energy functional of this form, and truncating at second order in the
gradients, the Euler-Lagrange equations, eqs.(\ref{dft13}), would become%
\begin{gather}
K_{ij}^{ab}\left( \Gamma \left( \overrightarrow{r}\right) \right) \frac{%
\partial ^{2}}{\partial r_{i}\partial r_{j}}\Gamma _{b}\left( 
\overrightarrow{r}\right) +g_{ij}^{abc}\left( \Gamma \left( \overrightarrow{r%
}\right) \right) \left( \frac{\partial }{\partial r_{i}}\Gamma _{b}\left( 
\overrightarrow{r}\right) \right) \left( \frac{\partial }{\partial r_{j}}%
\Gamma _{c}\left( \overrightarrow{r}\right) \right) -\frac{\partial }{%
\partial \Gamma _{a}\left( \overrightarrow{r}\right) }\beta F_{bulk}\left(
\Gamma \left( \overrightarrow{r}\right) \right) +\mu \delta _{a0}=0,
\label{EL2} \\
g_{ij}^{abc}\left( \Gamma \right) =\frac{\partial }{\partial \Gamma _{c}}%
K_{ij}^{ab}\left( \Gamma \right) -\frac{1}{2}\frac{\partial }{\partial
\Gamma _{a}}K_{ij}^{bc}\left( \Gamma \right)  \notag
\end{gather}%
where it is assumed that $\Gamma _{0}$ is the average density. These
variational equations are derived under the usual assumption that the values
of the order parameters on the boundaries are held fixed.

The idea presented here is to first break up the volume of the systems into
cells containing a single atom. Then, withing the cells, the spatial
dependence of the slowly varying parameters is expressed as a gradient
expansion. Since these parameters are, \textit{a priori} assumed to vary
little over atomic length scales, it is assumed that the gradient expansion
can be truncated without serious approximation. Finally, the resulting
expressions for each cell are resummed thus separating the atomic and
structural length scales.

\subsection{Initial gradient expansion}

To begin, it is convenient to use the representation%
\begin{equation}
\beta F\left[ \rho \right] =\int d\overrightarrow{r}\;\beta f\left( 
\overrightarrow{r};\left[ \rho \right] \right)  \label{e1}
\end{equation}%
where $f\left( \overrightarrow{r};\left[ \Gamma \right] \right) $ can be
expressed in terms of the one-body dcf and the density using the methods of
the previous Section,%
\begin{equation}
\beta f\left( \overrightarrow{r};\left[ \rho \right] \right)
=-\int_{0}^{1}\;c_{1}\left( \overrightarrow{r};\left[ \lambda \rho \right]
\right) \rho \left( \overrightarrow{r}\right) d\lambda .
\end{equation}%
An important assumption made throughout this Section is that the density is
parameterized by a set of order parameters which might include e.g. the
average density but that the underlying lattice structure, including the
lattice constant, is held fixed. For example, the Haymet-Oxtoby
parameterization given in eq.(\ref{p1}) is of this form since the reciprocal
lattice vectors $\overrightarrow{K}_{n}$ are held fixed; another example is
that used in ref.(\cite{OLW})%
\begin{equation}
\rho \left( \overrightarrow{r}\right) =\eta _{0}\left( \overrightarrow{r}%
\right) \left( \frac{\alpha \left( \overrightarrow{r}\right) }{\pi }\right)
^{3/2}\sum_{n}\exp \left( -\alpha \left( \overrightarrow{r}\right) \left( 
\overrightarrow{r}-\overrightarrow{R}_{n}\right) ^{2}\right)  \label{p2}
\end{equation}%
where the two order parameters are $\Gamma _{0}\left( \overrightarrow{r}%
\right) =\eta _{0}\left( \overrightarrow{r}\right) $ and $\Gamma _{1}\left( 
\overrightarrow{r}\right) =\alpha \left( \overrightarrow{r}\right) $ but the
lattice vectors $\overrightarrow{R}_{n}$ are held constant. Furthermore,
attention here will be restricted to the most important case of simple
Bravis lattices, such as FCC and BCC, with a single atom per unit cell.
Notice that this also includes as a special case the nonuniform liquid which
results in both parameterizations from taking $\Gamma _{n}\left( 
\overrightarrow{r}\right) =0$ for $n\geq 0$. These models are therefore
sufficiently flexible to describe the solid-liquid interface, as well as
e.g. the gas-liquid interface.

To begin,  partition space into the Wigner-Seitz cells where the center of the $n$
th cell, $V_{n}$, will be at $\overrightarrow{R}_{n}$. and all cells have
volume $V_{ws}.$ The expression for the free energy can therefore be written
as 
\begin{equation}
\beta F\left[ \rho \right] =\sum_{n=1}^{N_{V}}\int_{V_{n}}d\overrightarrow{r}%
\;\beta f\left( \overrightarrow{r};\left[ \rho \right] \right)  \label{e2}
\end{equation}%
In each cell, the density is expanded about that of a system with uniform order
parameter 
\begin{eqnarray}
\beta F\left[ \rho \right] &=&\sum_{n=1}^{N_{V}}\int_{V_{n}}d\overrightarrow{%
r}\;\beta f\left( \overrightarrow{r};\left[ \rho _{n}\right] \right)
\label{e3} \\
&&+\sum_{n=1}^{N_{V}}\int_{V_{n}}d\overrightarrow{r}\int d\overrightarrow{r}%
_{1}\;\left. \frac{\delta \beta f\left( \overrightarrow{r};\left[ \rho %
\right] \right) }{\delta \rho \left( \overrightarrow{r}_{1}\right) }\right|
_{\rho _{n}}\delta \rho \left( \overrightarrow{r}_{1};\Gamma \left( 
\overrightarrow{r}_{1}\right) ,\Gamma _{n}\right)  \notag \\
&&+\frac{1}{2}\sum_{n=1}^{N_{V}}\int_{V_{n}}d\overrightarrow{r}\int d%
\overrightarrow{r}_{1}d\overrightarrow{r}_{2}\;\left. \frac{\delta ^{2}\beta
f\left( \overrightarrow{r};\left[ \rho \right] \right) }{\delta \rho \left( 
\overrightarrow{r}_{1}\right) \delta \rho \left( \overrightarrow{r}%
_{2}\right) }\right| _{\rho _{n}}\delta \rho \left( \overrightarrow{r}%
_{1};\Gamma \left( \overrightarrow{r}_{1}\right) ,\Gamma _{n}\right) \delta
\rho \left( \overrightarrow{r}_{2};\Gamma \left( \overrightarrow{r}%
_{2}\right) ,\Gamma _{n}\right)  \notag \\
&&+...  \notag
\end{eqnarray}%
where%
\begin{equation}
\delta \rho \left( \overrightarrow{r}_{1};\Gamma \left( \overrightarrow{r}%
_{1}\right) ,\Gamma _{n}\right) =\rho \left( \overrightarrow{r}_{1};\Gamma
\left( \overrightarrow{r}_{1}\right) \right) -\rho \left( \overrightarrow{r}%
_{1};\Gamma _{n}\right)
\end{equation}%
and $\Gamma _{n}=\Gamma \left( \overrightarrow{R}_{n}\right) $ is the value
of order parameter at the center of the nth cell. At this point, all terms are retained and no truncation is imposed. Next, the dependence of the densities on the order parameters is expanded using%
\begin{eqnarray}
\rho \left( \overrightarrow{r}_{1};\Gamma \left( \overrightarrow{r}%
_{1}\right) \right) &=&\rho \left( \overrightarrow{r}_{1};\Gamma _{n}\right)
+\sum_{a}\left( \Gamma _{a}\left( \overrightarrow{r}_{1}\right) -\Gamma
_{n,a}\right) \frac{\partial \rho \left( \overrightarrow{r}_{1};\Gamma
_{n}\right) }{\partial \Gamma _{n,a}}  \label{e4} \\
&&+\frac{1}{2}\sum_{ab}\left( \Gamma _{a}\left( \overrightarrow{r}%
_{1}\right) -\Gamma _{n,a}\right) \left( \Gamma _{b}\left( \overrightarrow{r}%
_{1}\right) -\Gamma _{n,b}\right) \frac{\partial ^{2}\rho \left( 
\overrightarrow{r}_{1};\Gamma _{n}\right) }{\partial \Gamma _{n,a}\partial
\Gamma _{n,b}}  \notag \\
&&+...  \notag
\end{eqnarray}%
as is the spatial dependence of the order parameters themselves
\begin{equation}
\Gamma _{a}\left( \overrightarrow{r}_{1}\right) =\Gamma _{a}\left( 
\overrightarrow{R}_{n}\right) +\left( \overrightarrow{r}_{1}-\overrightarrow{%
R}_{n}\right) \cdot \left( \overrightarrow{\nabla }\Gamma _{a}\right) _{n}+%
\frac{1}{2}\left( \overrightarrow{r}_{1}-\overrightarrow{R}_{n}\right)
\left( \overrightarrow{r}_{1}-\overrightarrow{R}_{n}\right) :\left( 
\overrightarrow{\nabla }\overrightarrow{\nabla }\Gamma _{a}\right) _{n}+...
\label{e5}
\end{equation}%
The key assumption made is that terms of order $\nabla ^{n}\Gamma $ may be
neglected for some $n>n_{0}.$ Then, the functional expansion, eq.(\ref{e3}),
is only needed up to order $n_{0}$ as well. Taking $n_{0}=2$ and combining
eqs.(\ref{e2}-\ref{e5}) gives%
\begin{eqnarray}
\beta F\left[ \Gamma \right] &=&\sum_{n=0}^{N_{V}}\int_{V_{n}}d%
\overrightarrow{r}\;\beta f\left( \overrightarrow{r};\left[ \rho _{n}\right]
\right) \\
&&+\sum_{n=0}^{N_{V}}\int_{V_{n}}d\overrightarrow{r}\int d\overrightarrow{r}%
_{1}\;\left. \frac{\delta \beta f\left( \overrightarrow{r};\left[ \Gamma %
\right] \right) }{\delta \rho \left( \overrightarrow{r}_{1}\right) }\right|
_{\rho _{n}}\left( \overrightarrow{r}_{1}-\overrightarrow{R}_{n}\right)
\cdot \left( \overrightarrow{\nabla }\Gamma _{n,a}\right) \frac{\partial
\rho \left( \overrightarrow{r}_{1};\Gamma _{n}\right) }{\partial \Gamma
_{n,a}}  \notag \\
&&+\frac{1}{2}\sum_{n=0}^{N_{V}}\int_{V_{n}}d\overrightarrow{r}\int d%
\overrightarrow{r}_{1}\;\left. \frac{\delta \beta f\left( \overrightarrow{r};%
\left[ \Gamma \right] \right) }{\delta \rho \left( \overrightarrow{r}%
_{1}\right) }\right| _{\rho _{n}}\left( \overrightarrow{r}_{1}-%
\overrightarrow{R}_{n}\right) \left( \overrightarrow{r}_{1}-\overrightarrow{R%
}_{n}\right) :\left( \overrightarrow{\nabla }\overrightarrow{\nabla }\Gamma
_{n,a}\right) \frac{\partial \rho \left( \overrightarrow{r}_{1};\Gamma
_{n}\right) }{\partial \Gamma _{n,a}}  \notag \\
&&+\frac{1}{2}\sum_{n=0}^{N_{V}}\int_{V_{n}}d\overrightarrow{r}\int d%
\overrightarrow{r}_{1}\;\left. \frac{\delta \beta f\left( \overrightarrow{r};%
\left[ \Gamma \right] \right) }{\delta \rho \left( \overrightarrow{r}%
_{1}\right) }\right| _{\rho _{n}}\left( \overrightarrow{r}_{1}-%
\overrightarrow{R}_{n}\right) \left( \overrightarrow{r}_{1}-\overrightarrow{R%
}_{n}\right) :\left( \overrightarrow{\nabla }\Gamma _{n,a}\right) \left( 
\overrightarrow{\nabla }\Gamma _{n,b}\right) \frac{\partial ^{2}\rho \left( 
\overrightarrow{r}_{1};\Gamma _{n}\right) }{\partial \Gamma _{n,a}\partial
\Gamma _{n,b}}  \notag \\
&&+\frac{1}{2}\sum_{n=0}^{N_{V}}\int_{V_{n}}d\overrightarrow{r}\int d%
\overrightarrow{r}_{1}d\overrightarrow{r}_{2}\;\left. \frac{\delta ^{2}\beta
f\left( \overrightarrow{r};\left[ \Gamma \right] \right) }{\delta \rho
\left( \overrightarrow{r}_{1}\right) \delta \rho \left( \overrightarrow{r}%
_{2}\right) }\right| _{\rho _{n}}\left[ 
\begin{array}{c}
\left( \overrightarrow{r}_{1}-\overrightarrow{R}_{n}\right) \left( 
\overrightarrow{r}_{2}-\overrightarrow{R}_{n}\right) :\left( \overrightarrow{%
\nabla }\Gamma _{n,a}\right) \left( \overrightarrow{\nabla }\Gamma
_{n,b}\right) \\ 
\times \frac{\partial \rho \left( \overrightarrow{r}_{1};\Gamma _{n}\right) 
}{\partial \Gamma _{n,a}}\frac{\partial \rho \left( \overrightarrow{r}%
_{2};\Gamma _{n}\right) }{\partial \Gamma _{n,b}}%
\end{array}%
\right]  \notag
\end{eqnarray}%
Because the terms involving functional derivatives are being evaluated at
constant values of the order parameters, they exhibit the symmetry of a bulk
solid. In particular, they are invariant with respect to translations by a
lattice vector so a change of variables yields%
\begin{eqnarray}
\beta F\left[ \Gamma \right] &=&\sum_{n=0}^{N_{V}}\int_{V_{n}}d%
\overrightarrow{r}\;\beta f\left( \overrightarrow{r};\left[ \rho _{n}\right]
\right) \\
&&+\sum_{n=0}^{N_{V}}\int_{V_{n}}d\overrightarrow{r}\int d\overrightarrow{r}%
_{1}\;\left. \frac{\delta \beta f\left( \overrightarrow{r};\left[ \Gamma %
\right] \right) }{\delta \rho \left( \overrightarrow{r}_{1}\right) }\right|
_{\rho _{n}}\overrightarrow{r}_{1}\cdot \left( \overrightarrow{\nabla }%
\Gamma _{n,a}\right) \frac{\partial \rho \left( \overrightarrow{r}%
_{1};\Gamma _{n}\right) }{\partial \Gamma _{n,a}}  \notag \\
&&+\frac{1}{2}\sum_{n=0}^{N_{V}}\int_{V_{n}}d\overrightarrow{r}\int d%
\overrightarrow{r}_{1}\;\left. \frac{\delta \beta f\left( \overrightarrow{r};%
\left[ \Gamma \right] \right) }{\delta \rho \left( \overrightarrow{r}%
_{1}\right) }\right| _{\rho _{n}}\overrightarrow{r}_{1}\overrightarrow{r}%
_{1}\cdot \left( \overrightarrow{\nabla }\overrightarrow{\nabla }\Gamma
_{n,a}\right) \frac{\partial \rho \left( \overrightarrow{r}_{1};\Gamma
_{n}\right) }{\partial \Gamma _{n,a}}  \notag \\
&&+\frac{1}{2}\sum_{n=0}^{N_{V}}\int_{V_{n}}d\overrightarrow{r}\int d%
\overrightarrow{r}_{1}\;\left. \frac{\delta \beta f\left( \overrightarrow{r};%
\left[ \Gamma \right] \right) }{\delta \rho \left( \overrightarrow{r}%
_{1}\right) }\right| _{\rho _{n}}\overrightarrow{r}_{1}\overrightarrow{r}%
_{1}\cdot \left( \overrightarrow{\nabla }\Gamma _{n,a}\right) \left( 
\overrightarrow{\nabla }\Gamma _{n,b}\right) \frac{\partial ^{2}\rho \left( 
\overrightarrow{r}_{1};\Gamma _{n}\right) }{\partial \Gamma _{n,a}\partial
\Gamma _{n,b}}  \notag \\
&&+\frac{1}{2}\sum_{n=0}^{N_{V}}\int_{V_{n}}d\overrightarrow{r}\int d%
\overrightarrow{r}_{1}d\overrightarrow{r}_{2}\;\left. \frac{\delta ^{2}\beta
f\left( \overrightarrow{r};\left[ \Gamma \right] \right) }{\delta \rho
\left( \overrightarrow{r}_{1}\right) \delta \rho \left( \overrightarrow{r}%
_{2}\right) }\right| _{\rho _{n}}\overrightarrow{r}_{1}\overrightarrow{r}%
_{2}\cdot \left( \overrightarrow{\nabla }\Gamma _{n,a}\right) \left( 
\overrightarrow{\nabla }\Gamma _{n,b}\right) \frac{\partial \rho \left( 
\overrightarrow{r}_{1};\Gamma _{n}\right) }{\partial \Gamma _{n,a}}\frac{%
\partial \rho \left( \overrightarrow{r}_{2};\Gamma _{n}\right) }{\partial
\Gamma _{n,b}}  \notag
\end{eqnarray}%
The second term on the right vanishes by virtue of reflection symmetry in a
Bravais lattice, $\rho \left( -\overrightarrow{r}_{1};\Gamma _{n}\right)
=\rho \left( \overrightarrow{r}_{1};\Gamma _{n}\right) $. The only explicit
dependence on $\overrightarrow{r}$ is via $\phi \left( \overrightarrow{r};%
\left[ \rho _{n}\right] \right) $ and the integral of this quantity and its
derivatives over $V_{0}$ can be extended to an integral over all space and
using 
\begin{eqnarray}
\int_{V_{0}}d\overrightarrow{r}\;f\left( \overrightarrow{r};\left[ \rho _{n}%
\right] \right) &=&\frac{1}{N}\int d\overrightarrow{r}\;f\left( 
\overrightarrow{r};\left[ \rho _{n}\right] \right) =\frac{1}{N}F_{ex}\left[
\rho _{n}\right] \\
\int_{V_{n}}d\overrightarrow{r}\;\left. \frac{\delta \beta f\left( 
\overrightarrow{r};\left[ \Gamma \right] \right) }{\delta \rho \left( 
\overrightarrow{r}_{1}\right) }\right| _{\rho _{n}} &=&\frac{1}{N}\left. 
\frac{\delta \beta F_{ex}\left[ \rho \right] }{\delta \rho \left( 
\overrightarrow{r}_{1}\right) }\right| _{\rho _{n}}=-\frac{1}{N}c_{1}\left( 
\overrightarrow{r}_{1};\left[ \rho _{n}\right] \right)  \notag \\
\int_{V_{n}}d\overrightarrow{r}\;\left. \frac{\delta ^{2}f\left( 
\overrightarrow{r};\left[ \Gamma \right] \right) }{\delta \rho \left( 
\overrightarrow{r}_{1}\right) \delta \rho \left( \overrightarrow{r}%
_{2}\right) }\right| _{\rho _{n}} &=&\frac{1}{N}\left. \frac{\delta
^{2}\beta F_{ex}\left[ \rho \right] }{\delta \rho \left( \overrightarrow{r}%
_{1}\right) \delta \rho \left( \overrightarrow{r}_{2}\right) }\right| _{\rho
_{n}}=-\frac{1}{N}c_{2}\left( \overrightarrow{r}_{1},\overrightarrow{r}_{2};%
\left[ \rho _{n}\right] \right)  \notag
\end{eqnarray}%
gives%
\begin{eqnarray}
\beta F\left[ \Gamma \right] &=&\frac{1}{N}\sum_{n=0}^{N_{V}}F_{ex}\left[
\rho _{n}\right] \\
&&-\frac{1}{2N}\sum_{n=0}^{N_{V}}\int d\overrightarrow{r}_{1}\;c_{1}\left( 
\overrightarrow{r}_{1};\left[ \rho _{n}\right] \right) \overrightarrow{r}_{1}%
\overrightarrow{r}_{1}\cdot \left( \overrightarrow{\nabla }\overrightarrow{%
\nabla }\Gamma _{n,a}\right) \frac{\partial \rho \left( \overrightarrow{r}%
_{1};\Gamma _{n}\right) }{\partial \Gamma _{n,a}}  \notag \\
&&-\frac{1}{2N}\sum_{n=0}^{N_{V}}\int d\overrightarrow{r}_{1}\;c_{1}\left( 
\overrightarrow{r}_{1};\left[ \rho _{n}\right] \right) \overrightarrow{r}_{1}%
\overrightarrow{r}_{1}\cdot \left( \overrightarrow{\nabla }\Gamma
_{n,a}\right) \left( \overrightarrow{\nabla }\Gamma _{n,b}\right) \frac{%
\partial ^{2}\rho \left( \overrightarrow{r}_{1};\Gamma _{n}\right) }{%
\partial \Gamma _{n,a}\partial \Gamma _{n,b}}  \notag \\
&&-\frac{1}{2N}\sum_{n=0}^{N_{V}}\int d\overrightarrow{r}_{1}d%
\overrightarrow{r}_{2}\;c_{2}\left( \overrightarrow{r}_{1},\overrightarrow{r}%
_{2};\left[ \rho _{n}\right] \right) \overrightarrow{r}_{1}\overrightarrow{r}%
_{2}\cdot \left( \overrightarrow{\nabla }\Gamma _{n,a}\right) \left( 
\overrightarrow{\nabla }\Gamma _{n,b}\right) \frac{\partial \rho \left( 
\overrightarrow{r}_{1};\Gamma _{n}\right) }{\partial \Gamma _{n,a}}\frac{%
\partial \rho \left( \overrightarrow{r}_{2};\Gamma _{n}\right) }{\partial
\Gamma _{n,b}}  \notag
\end{eqnarray}%
This represents the desired coarse-graining of the dependence of the order
parameter on the microscopic structure. This is made clearer by writing the
free energy as%
\begin{equation}
\beta F\left[ \Gamma \right] =\sum_{n=0}^{N_{V}}\left( 
\begin{array}{c}
K_{0}\left( \Gamma \left( \overrightarrow{R}_{n}\right) \right) +\sum_{a}%
\overleftrightarrow{K}_{1,a}\left( \Gamma \left( \overrightarrow{R}%
_{n}\right) \right) :\left( \overrightarrow{\nabla }\overrightarrow{\nabla }%
\Gamma _{a}\right) _{\overrightarrow{r}=\overrightarrow{R}_{n}} \\ 
+\sum_{ab}\overleftrightarrow{K}_{2,ab}\left( \Gamma \left( \overrightarrow{R%
}_{n}\right) \right) :\left( \overrightarrow{\nabla }\Gamma _{a}\right) _{%
\overrightarrow{r}=\overrightarrow{R}_{n}}\left( \overrightarrow{\nabla }%
\Gamma _{b}\right) _{\overrightarrow{r}=\overrightarrow{R}_{n}}%
\end{array}%
\right)  \label{separated}
\end{equation}%
with%
\begin{eqnarray}
K_{0}\left( \Gamma \left( \overrightarrow{R}_{n}\right) \right) &=&\frac{1}{N%
}F_{ex}\left[ \rho _{n}\right] \\
\overleftrightarrow{K}_{1,a}\left( \Gamma \left( \overrightarrow{R}%
_{n}\right) \right) &=&-\frac{1}{2N}\int d\overrightarrow{r}%
_{1}\;c_{1}\left( \overrightarrow{r}_{1};\left[ \rho _{n}\right] \right) 
\overrightarrow{r}_{1}\overrightarrow{r}_{1}\frac{\partial \rho \left( 
\overrightarrow{r}_{1};\Gamma _{n}\right) }{\partial \Gamma _{n,a}}  \notag
\\
\overleftrightarrow{K}_{2,ab}\left( \Gamma \left( \overrightarrow{R}%
_{n}\right) \right) &=&-\frac{1}{2N}\int d\overrightarrow{r}%
_{1}\;c_{1}\left( \overrightarrow{r}_{1};\left[ \rho _{n}\right] \right) 
\overrightarrow{r}_{1}\overrightarrow{r}_{1}\frac{\partial ^{2}\rho \left( 
\overrightarrow{r}_{1};\Gamma _{n}\right) }{\partial \Gamma _{n,a}\partial
\Gamma _{n,b}}  \notag \\
&&-\frac{1}{2N}\int d\overrightarrow{r}_{1}d\overrightarrow{r}%
_{2}\;c_{2}\left( \overrightarrow{r}_{1},\overrightarrow{r}_{2};\left[ \rho
_{n}\right] \right) \overrightarrow{r}_{1}\overrightarrow{r}_{2}\frac{%
\partial \rho \left( \overrightarrow{r}_{1};\Gamma _{n}\right) }{\partial
\Gamma _{n,a}}\frac{\partial \rho \left( \overrightarrow{r}_{2};\Gamma
_{n}\right) }{\partial \Gamma _{n,b}}  \notag
\end{eqnarray}

\subsection{Resummation}

Having obtained the form given in eq.(\ref{separated}), the question is
whether the lattice sums might be written as integrals so as to obtain
something like the postulated form given in eq.(\ref{Target}). This question
has not been addressed in previous derivations. To do the resummation, we
first write the lattice sums more explicitly. Let the primitive lattice
vectors for the $D$-dimensional lattice be $\left\{ \overrightarrow{a}%
_{i}\right\} _{i=1}^{D}$ so that the set of lattice vectors can be written
as $\left\{ \overrightarrow{R}_{n}\right\} _{n=1}^{n_{V}}=\left\{
\sum_{i=1}^{D}m_{i}\overrightarrow{a}_{i}\right\}
_{m_{1}...m_{D}=M_{1}...M_{D}}^{N_{1}...N_{D}}$ where the various limits, $%
M_{i}$ and $N_{i}$, will depend on the geometry. Then, for any function of
position $f\left( \overrightarrow{r}\right) $ one has that%
\begin{equation}
\sum_{n}f\left( \overrightarrow{R}_{n}\right)
=\sum_{m_{D}=M_{D}}^{N_{D}}...\sum_{m_{1}=M_{1}\left( m_{2}...m_{D}\right)
}^{N_{1}\left( m_{2}...m_{D}\right) }f\left( m_{1}\overrightarrow{a}%
_{1}+...+m_{D}\overrightarrow{a}_{D}\right) ,
\end{equation}%
where, as indicated, the limits of the inner sums  must be allowed to depend on
the indices of the outer sums so as to allow for the most general boundary
conditions. For simplicity, the dependence of the limits on
the indices will be suppressed but it should always be considered to be present. This summation
can be expressed as an integral by means of the Euler-Maclaurin summation
formula giving the exact relation%
\begin{eqnarray}
&&\sum_{m_{1}=M_{1}}^{N_{1}}f\left( m_{1}\overrightarrow{a}_{1}+...+m_{D}%
\overrightarrow{a}_{D}\right)  \label{Euler-Maclaurin} \\
&=&\int_{M_{1}}^{N_{1}}\left[ f\left( x\overrightarrow{a}_{1}+...+m_{D}%
\overrightarrow{a}_{D}\right) +\frac{\left( -1\right) ^{r}}{(r+1)!}%
B_{r+1}\left( x\right) \frac{\partial ^{r+1}}{\partial x^{r+1}}f\left( x%
\overrightarrow{a}_{1}+...+m_{D}\overrightarrow{a}_{D}\right) \right] dx 
\notag \\
&&+\sum_{k=1}^{r}\frac{\left( -1\right) ^{k+1}B_{k+1}}{(k+1)!}\left[ \frac{%
\partial ^{k}}{\partial x^{k}}f\left( x\overrightarrow{a}_{1}+...+m_{D}%
\overrightarrow{a}_{D}\right) \right] _{M_{1}}^{N_{1}}  \notag
\end{eqnarray}%
where $B_{k}$ is the k-th Bernoulli number, $B_{k}\left( x\right) $ is the
k-th Bernoulli periodic function and $r$ is an arbitrarily chosen positive
integer. The Bernoulli functions are polynomials for $x\in \lbrack 0,1]$ and
are defined outside this interval by $B_{k}\left( x\right) =B_{k}\left( x%
\text{mod}1\right) $ . Thus, they are bounded for all values of $k$ and in
fact it is also true that their integral over the unit interval vanishes.
The idea here is that since only terms up to order $n_{0}$ are retained, and
since the functions $B_{k}\left( x\right) $ are bounded, the last term on
the right can be neglected provided $r+1>n_{0}$. The second term on the
right is evaluated on the boundary. For $r=1$ , it vanishes provided 
\begin{equation}
\left. \frac{\partial }{\partial x}f\left( x\overrightarrow{a}_{1}+...+m_{D}%
\overrightarrow{a}_{D}\right) \right| _{x=N_{1}}=\left. \frac{\partial }{%
\partial x}f\left( x\overrightarrow{a}_{1}+...+m_{D}\overrightarrow{a}%
_{D}\right) \right| _{x=M_{1}}  \label{bc}
\end{equation}%
which could be true either because of the boundary conditions or because of
symmetry. Even if it does not vanish, its contribution will be of order $%
\frac{1}{L}$ compared to the first term , where $L=\left| \left(
N_{1}-M_{1}\right) \overrightarrow{a}_{1}\right| $ so that in the
thermodynamic limit, it should not contribute. Choosing $r$ to be
arbitrarily large and demanding that 
\begin{equation}
\sum_{k=1}^{r}\frac{\left( -1\right) ^{k+1}B_{k+1}}{(k+1)!}\left[ \frac{%
\partial ^{k}}{\partial x^{k}}f\left( x\overrightarrow{a}_{1}+...+m_{D}%
\overrightarrow{a}_{D}\right) \right] _{M_{1}}^{N_{1}}=0
\end{equation}%
imposes a more complex boundary condition. To illustrate, in one dimension, $%
r=3$ gives%
\begin{equation}
\frac{a_{1}}{12}\left[ f^{\prime }\left( N_{1}a_{1}\right) -f^{\prime
}\left( M_{1}a_{1}\right) \right] -\frac{a_{1}^{3}}{720}\left[ f^{(3)}\left(
N_{1}a_{1}\right) -f^{(3)}\left( M_{1}a_{1}\right) \right] =0.
\end{equation}%
It could be argued that in the present case, since only terms
up to second order in the gradients of the order parameters are kept, the third order
terms in this boundary condition could in any case be neglected so that any
choice of $r\geq 2$ would be satisfactory. It seems best, however, to note
that $B_{3}=0$ so that taking $r=2$ avoids any subtleties that might arise
on the boundary. Similarly, if one were keeping gradients up to order $n_{0},$
and $n_{0}$ is even, one would take $r=n_{0}$ with the consequent boundary
conditions involving derives up to order $n_{0}-1$. No such simple choice is
available if $n_{0}$ is odd.

So, for the present case of a second order theory taking $r=2$ gives,
when third derivatives are neglected,%
\begin{equation}
\sum_{m_{1}=M_{1}}^{N_{1}}f\left( m_{1}\overrightarrow{a}_{1}+...+m_{D}%
\overrightarrow{a}_{D}\right) \simeq \int_{M_{1}}^{N_{1}}f\left( x%
\overrightarrow{a}_{1}+...+m_{D}\overrightarrow{a}_{D}\right) dx,
\end{equation}%
together with the boundary condition, eq.(\ref{bc}). Furthermore, it happens
that $\left| B_{3}\left( x\right) \right| \leq $ $\frac{1}{36}\sqrt{3}\simeq
\allowbreak 5\times 10^{-2}$ and that $B_{3}\left( x\right) $ is oscillatory
over the distance of one lattice spacing whereas the order parameters are
meant to vary slowly over such small distances, thus further strengthening
the argument for neglecting the higher-order term. When this is extended to
include the second sum, a new complication arises since this gives%
\begin{eqnarray}
\sum_{m_{2}=M_{2}}^{N_{2}}\sum_{m_{1}=M_{1}}^{N_{1}}f\left( m_{1}%
\overrightarrow{a}_{1}+...+m_{D}\overrightarrow{a}_{D}\right) &\simeq
&\int_{M_{2}}^{N_{2}}dy\;\int_{M_{1}}^{N_{1}}dx\;f\left( x\overrightarrow{a}%
_{1}+y\overrightarrow{a}_{2}+...+m_{D}\overrightarrow{a}_{D}\right)
\label{resum2} \\
&&+\frac{B_{2}}{2}\left[ \frac{\partial }{\partial y}\int_{M_{1}}^{N_{1}}dx%
\;f\left( x\overrightarrow{a}_{1}+y\overrightarrow{a}_{2}...+m_{D}%
\overrightarrow{a}_{D}\right) \right] _{M_{2}}^{N_{2}}  \notag \\
&&+\frac{1}{3!}\int_{M_{2}}^{N_{2}}dyB_{3}\left( y\right) \frac{\partial ^{3}%
}{\partial y^{3}}\int_{M_{1}}^{N_{1}}dx\;f\left( x\overrightarrow{a}_{1}+y%
\overrightarrow{a}_{2}+...+m_{D}\overrightarrow{a}_{D}\right)  \notag
\end{eqnarray}%
In the event that $M_{1}$ and $N_{1}$ depend on $y$, as they would e.g. for
a spherical boundary, then the second term on the right becomes 
\begin{eqnarray}
&&\int_{M_{1}}^{N_{1}}dx\;\left[ \frac{\partial }{\partial y}\;f\left( x%
\overrightarrow{a}_{1}+y\overrightarrow{a}_{2}...+m_{D}\overrightarrow{a}%
_{D}\right) \right] _{M_{2}}^{N_{2}} \\
&&+\left[ \frac{\partial N_{1}\left( y\right) }{\partial y}\;f\left( N_{1}%
\overrightarrow{a}_{1}+y\overrightarrow{a}_{2}...+m_{D}\overrightarrow{a}%
_{D}\right) -\frac{\partial M_{1}}{\partial y}f\left( M_{1}\overrightarrow{a}%
_{1}+y\overrightarrow{a}_{2}...+m_{D}\overrightarrow{a}_{D}\right) \right]
_{M_{2}}^{N_{2}}.  \notag
\end{eqnarray}%
Now, the first contribution can be required to vanish giving a simple and
expected multidimensional generalization of eq.(\ref{bc}) while the second
term involves only the value of $f$ on the boundary and geometric functions
concerning the definition of the boundary. All of these quantities are fixed
in the variational procedure leading from eq.(\ref{Target}) to the
Euler-Lagrange equations, eq.(\ref{EL2}), so they are of no importance.
Thus, with the necessary generalization of the boundary condition, the
second term on the right of eq.(\ref{resum2}) can be neglected. Finally, the third
term on the right in eq.(\ref{resum2}) becomes 
\begin{eqnarray}
&&\int_{M_{2}}^{N_{2}}dyB_{3}\left( y\right) \left[ \frac{\partial ^{3}N_{1}%
}{\partial y^{3}}f\left( N_{1}\overrightarrow{a}_{1}+y\overrightarrow{a}%
_{2}+...+m_{D}\overrightarrow{a}_{D}\right) -\frac{\partial ^{3}M_{1}}{%
\partial y^{3}}f\left( M_{1}\overrightarrow{a}_{1}+y\overrightarrow{a}%
_{2}+...+m_{D}\overrightarrow{a}_{D}\right) \right] \\
&&+3\int_{M_{2}}^{N_{2}}dyB_{3}\left( y\right) \left[ \frac{\partial
^{2}N_{1}}{\partial y^{2}}\frac{\partial }{\partial y}f\left( N_{1}%
\overrightarrow{a}_{1}+y\overrightarrow{a}_{2}+...+m_{D}\overrightarrow{a}%
_{D}\right) -\frac{\partial ^{2}M_{1}}{\partial y^{2}}\frac{\partial }{%
\partial y}f\left( M_{1}\overrightarrow{a}_{1}+y\overrightarrow{a}%
_{2}+...+m_{D}\overrightarrow{a}_{D}\right) \right]  \notag \\
&&+3\int_{M_{2}}^{N_{2}}dyB_{3}\left( y\right) \left[ \frac{\partial N_{1}}{%
\partial y}\frac{\partial ^{2}}{\partial y^{2}}f\left( N_{1}\overrightarrow{a%
}_{1}+y\overrightarrow{a}_{2}+...+m_{D}\overrightarrow{a}_{D}\right) -\frac{%
\partial M_{1}}{\partial y}\frac{\partial ^{2}}{\partial y^{2}}f\left( M_{1}%
\overrightarrow{a}_{1}+y\overrightarrow{a}_{2}+...+m_{D}\overrightarrow{a}%
_{D}\right) \right]  \notag \\
&&+\int_{M_{2}}^{N_{2}}dyB_{3}\left( y\right) \int_{M_{1}}^{N_{1}}dx\;\frac{%
\partial ^{3}}{\partial y^{3}}f\left( x\overrightarrow{a}_{1}+y%
\overrightarrow{a}_{2}+...+m_{D}\overrightarrow{a}_{D}\right) .  \notag
\end{eqnarray}%
Here, the first term is of no importance since it involves the value of $f$
on the boundary and the fourth term can be neglected as it is of third
order. Note that in the remaining terms, $\frac{\partial }{\partial y}%
f\left( N_{1}\overrightarrow{a}_{1}+y\overrightarrow{a}_{2}+...+m_{D}%
\overrightarrow{a}_{D}\right) $ is a derivative evaluated for variations on
the boundary, as are all of the other derivatives that occur in the second
and third terms. Thus, if the boundary condition is that the order
parameters assume constant values on the boundary, then these terms are
negligible. This type of boundary condition is certainly reasonable for
spherically symmetric problems or on the radial boundary for those with
cylindrical symmetry. On the other hand, for planar interfaces one could
take the volume to be defined by $N_{i}=-M_{i}=N$ for some constant $N$ in
which case $\frac{\partial N_{1}}{\partial y}=0$, etc. and again the second
and third contributions are negligible. Again, these terms should be of no
consequence in the thermodynamic limit as they are of order $\frac{1}{L}$
compared to the first term in eq.(\ref{resum2}). Thus, with these
considerations, it has been proven that up to third order gradients, 
\begin{equation}
\sum_{m_{2}=M_{2}}^{N_{2}}\sum_{m_{1}=M_{1}}^{N_{1}}f\left( m_{1}%
\overrightarrow{a}_{1}+...+m_{D}\overrightarrow{a}_{D}\right) \simeq
\int_{M_{2}}^{N_{2}}dy\;\int_{M_{1}}^{N_{1}}dx\;f\left( x\overrightarrow{a}%
_{1}+y\overrightarrow{a}_{2}+...+m_{D}\overrightarrow{a}_{D}\right) .
\end{equation}%
Clearly, this result can be extended to any number of dimensions to get the
general result 
\begin{equation}
\sum_{m_{D}=M_{D}}^{N_{D}}...\sum_{m_{1}=N_{1}}^{N_{1}}f\left( m_{1}%
\overrightarrow{a}_{1}+...+m_{D}\overrightarrow{a}_{D}\right)
=\int_{M_{D}}^{N_{D}}...\int_{M_{1}}^{N_{1}}dx_{1}...dx_{D}\;f\left( x_{1}%
\overrightarrow{a}_{1}+...+x_{D}\overrightarrow{a}_{D}\right)
\end{equation}%
which is expected to hold as the limits go to infinity or, for finite systems,
with the corresponding boundary conditions%
\begin{equation}
\forall i,\left. \frac{\partial }{\partial x_{i}}f\left( x_{1}%
\overrightarrow{a}_{1}+...+x_{D}\overrightarrow{a}_{D}\right) \right|
_{x_{i}=N_{i}}=\left. \frac{\partial }{\partial x_{i}}f\left( x_{1}%
\overrightarrow{a}_{1}+...+x_{D}\overrightarrow{a}_{D}\right) \right|
_{x_{i}=M_{i}}.
\end{equation}%
and $f$ is either a constant on the boundaries or the boundaries are defined
by fixed values of $M_{i}$ and $N_{i}$. Since the primitive lattice vectors
are linearly independent, the integral can be written as a volume integral
by a change of variables giving%
\begin{equation}
\sum_{m_{D}=M_{D}}^{N_{D}}...\sum_{m_{1}=N_{1}}^{N_{1}}f\left( m_{1}%
\overrightarrow{a}_{1}+...+m_{D}\overrightarrow{a}_{D}\right) =\frac{1}{%
V_{ws}}\int_{V}\left[ f\left( \overrightarrow{r}\right) +O\left( \nabla
^{3}f\right) \right] d\overrightarrow{r}+\chi ,  \label{summation}
\end{equation}%
where $\chi $ represents the neglected terms which involve the value of $f$
on the boundaries and the various geometric factors. It is important in
terms of calculating the absolute value of the free energy, but plays no
role in formulating the Euler-Lagrange equations for the order parameters.

\subsection{Continuum Limit}

Making use of eq.(\ref{summation}) and $NV_{ws}=V$, one has that, up to
second order in the gradients and an overall additive constant, that%
\begin{eqnarray}
\beta F\left[ \Gamma \right] &=&\frac{1}{V}\int d\overrightarrow{R}\;\beta F%
\left[ \rho \left( \Gamma \left( \overrightarrow{R}\right) \right) \right] \\
&&-\frac{1}{2V}\int d\overrightarrow{r}_{1}d\overrightarrow{R}\;c_{1}\left( 
\overrightarrow{r}_{1};\left[ \rho \left( \Gamma \left( \overrightarrow{R}%
\right) \right) \right] \right) \overrightarrow{r}_{1}\overrightarrow{r}_{1}:%
\frac{\partial }{\partial \overrightarrow{R}}\left( \frac{\partial \Gamma
_{a}\left( \overrightarrow{R}\right) }{\partial \overrightarrow{R}}\frac{%
\partial \rho \left( \overrightarrow{r}_{1};\Gamma \left( \overrightarrow{R}%
\right) \right) }{\partial \Gamma _{a}\left( \overrightarrow{R}\right) }%
\right)  \notag \\
&&-\frac{1}{2V}\int d\overrightarrow{r}_{1}d\overrightarrow{r}_{2}d%
\overrightarrow{R}\;c_{12}\left( \overrightarrow{r}_{1},\overrightarrow{r}%
_{2};\left[ \rho \left( \Gamma \left( \overrightarrow{R}\right) \right) %
\right] \right) \left[ 
\begin{array}{c}
\overrightarrow{r}_{1}\overrightarrow{r}_{2}:\frac{\partial \Gamma
_{a}\left( \overrightarrow{R}\right) }{\partial \overrightarrow{R}}\frac{%
\partial \Gamma _{b}\left( \overrightarrow{R}\right) }{\partial 
\overrightarrow{R}} \\ 
\times \frac{\partial \rho \left( \overrightarrow{r}_{1};\Gamma \left( 
\overrightarrow{R}\right) \right) }{\partial \Gamma _{a}\left( 
\overrightarrow{R}\right) }\frac{\partial \rho \left( \overrightarrow{r}%
_{2};\Gamma \left( \overrightarrow{R}\right) \right) }{\partial \Gamma
_{a}\left( \overrightarrow{R}\right) }%
\end{array}%
\right] .  \notag
\end{eqnarray}%
An integration by parts gives%
\begin{eqnarray}
\beta F\left[ \Gamma \right] &=&\frac{1}{V}\int d\overrightarrow{R}\;\beta F%
\left[ \rho \left( \Gamma \left( \overrightarrow{R}\right) \right) \right] \\
&&-\frac{1}{2V}\int d\overrightarrow{r}_{1}d\overrightarrow{r}_{2}d%
\overrightarrow{R}\;\left[ 
\begin{array}{c}
c_{12}\left( \overrightarrow{r}_{1},\overrightarrow{r}_{2};\left[ \rho
\left( \Gamma \left( \overrightarrow{R}\right) \right) \right] \right)
\left( \overrightarrow{r}_{1}\overrightarrow{r}_{2}-\overrightarrow{r}_{1}%
\overrightarrow{r}_{1}\right) :\frac{\partial \Gamma _{a}\left( 
\overrightarrow{R}\right) }{\partial \overrightarrow{R}}\frac{\partial
\Gamma _{b}\left( \overrightarrow{R}\right) }{\partial \overrightarrow{R}}
\\ 
\times \frac{\partial \rho \left( \overrightarrow{r}_{1};\Gamma \left( 
\overrightarrow{R}\right) \right) }{\partial \Gamma _{a}\left( 
\overrightarrow{R}\right) }\frac{\partial \rho \left( \overrightarrow{r}%
_{2};\Gamma \left( \overrightarrow{R}\right) \right) }{\partial \Gamma
_{b}\left( \overrightarrow{R}\right) }%
\end{array}%
\right]  \notag \\
&&-\frac{1}{2V}\int d\overrightarrow{r}_{1}d\overrightarrow{R}\;\frac{%
\partial }{\partial \overrightarrow{R}}\cdot c_{1}\left( \overrightarrow{r}%
_{1};\left[ \rho \left( \Gamma \left( \overrightarrow{R}\right) \right) %
\right] \right) \overrightarrow{r}_{1}\overrightarrow{r}_{1}\cdot \left( 
\frac{\partial \Gamma _{a}\left( \overrightarrow{R}\right) }{\partial 
\overrightarrow{R}}\frac{\partial \rho \left( \overrightarrow{r}_{1};\Gamma
\left( \overrightarrow{R}\right) \right) }{\partial \Gamma _{a}\left( 
\overrightarrow{R}\right) }\right)
\end{eqnarray}%
or, rearranging the second term and using Gauss's theorem on the third term,%
\begin{eqnarray}
\beta F\left[ \Gamma \right] &=&\int d\overrightarrow{R}\;\left[ \frac{1}{V}%
\beta F\left( \Gamma \left( \overrightarrow{R}\right) \right) +\frac{1}{2}%
K_{ij}^{ab}\left( \Gamma \left( \overrightarrow{R}\right) \right) \frac{%
\partial \Gamma _{a}\left( \overrightarrow{R}\right) }{\partial R_{i}}\frac{%
\partial \Gamma _{b}\left( \overrightarrow{R}\right) }{\partial R_{j}}\right]
\label{Final} \\
&&-\frac{1}{2V}\int d\overrightarrow{r}_{1}\int_{S\left(
V\right) }c_{1}\left( \overrightarrow{r}_{1};\left[ \rho \left( \Gamma
\left( \overrightarrow{R}\right) \right) \right] \right) \overrightarrow{r}%
_{1}\cdot \left( \frac{\partial \Gamma _{a}\left( \overrightarrow{R}\right) 
}{\partial \overrightarrow{R}}\frac{\partial \rho \left( \overrightarrow{r}%
_{1};\Gamma \left( \overrightarrow{R}\right) \right) }{\partial \Gamma
_{a}\left( \overrightarrow{R}\right) }\right) \overrightarrow{r}_{1}\cdot d%
\overrightarrow{S}  \notag
\end{eqnarray}%
with%
\begin{equation}
K_{ij}^{ab}\left( \Gamma \right) =\frac{1}{2V}\int d\overrightarrow{r}_{1}d%
\overrightarrow{r}_{2}\;r_{12i}r_{12j}c_{2}\left( \overrightarrow{r}_{1},%
\overrightarrow{r}_{2};\rho _{\Gamma }\right) \frac{\partial \rho \left( 
\overrightarrow{r}_{1};\Gamma \right) }{\partial \Gamma _{a}}\frac{\partial
\rho \left( \overrightarrow{r}_{2};\Gamma \right) }{\partial \Gamma _{b}}.
\end{equation}%
and where the surface term involves an integral of $\overrightarrow{R}$ over
the boundary of the integration volume $V$ , $S(V)$. Except for the surface
term, this is the desired result. In the thermodynamic limit, the surface
term gives negligible contribution, as long as the order parameters and
their derivatives are finite on the boundary.

\section{Conclusions}

In this paper, it has been shown that, up to second order in the gradients
of the order parameters and neglecting various constants which depend on the
value of the density on the boundaries of the region of interest, the
Helmholtz free energy of an arbitrary simple crystalline system with fixed
lattice structure can be written in the form of eq.(\ref{Target}) thus
generalizing the results of L\"{o}wen, et al\cite{Lowen1},\cite{Lowen2}.
This expression makes a cleaner separations of length scales than does the
older form, eq.(\ref{EOH}). Furthermore, at no point in this derivation were
higher order correlations arbitrarily neglected: the only approximation made
was to truncate the gradient expansion at second order. This directly
addresses some criticisms of the use of these types of models, namely that
they are too crude in the treatment of correlations\cite%
{DftGradientCriticism}.

The applicability of eq.(\ref{Final}) depends on the boundary conditions. It
requires no qualification in the thermodynamic limit, except that it is
assumed that the order parameters and their derivatives are finite on the
boundary and the usual derivation of the Euler-Lagrange equations require
that the values of the order parameters be constant on the boundary. For
finite systems, the resummation of the local free energies requires that 
\begin{equation}
\forall i,\left. \frac{\partial }{\partial x_{i}}\Gamma \left( x_{1}%
\overrightarrow{a}_{1}+...+x_{D}\overrightarrow{a}_{D}\right) \right|
_{x_{i}=N_{i}}=\left. \frac{\partial }{\partial x_{i}}\Gamma \left( x_{1}%
\overrightarrow{a}_{1}+...+x_{D}\overrightarrow{a}_{D}\right) \right|
_{x_{i}=M_{i}}  \label{bc2}
\end{equation}%
and that $\Gamma $ is either a constant on the boundaries or that $\frac{%
\partial N_{i}}{\partial x_{j}}=0$ , etc.Problems involving spherical or
cylindrical symmetry will generally automatically ensure the constancy of $%
\Gamma $ on the boundary and planar interfaces can be handled by assuming a
volume defined by $N_{i}=-M_{i}=N$ for some constant $N$. For finite
systems, the surface term in eq.(\ref{Final}) must also be considered.

One noteworthy point about the Ginzburg-Landau form, eq.(\ref{Final}), is that
all of the microscopic quantities are evaluated for spatially uniform order
parameters or, in other words, for bulk systems. Thus, subject to the
approximation inherent in the gradient expansion, this technique provides a
framework for using good models of bulk systems to study interfacial systems.

One important problem not addressed here is that of elastic relaxation. In
general, one would like to allow the lattice parameter to be included in the
list of order parameters so that the lattice can expand or contract as one
passes from, say, a bulk solid phase into a bulk liquid phase. In fact, this
model has been used for a similar problem in which the lattice is allowed to
vary from FCC to BCC\cite{OxtobyBccFcc}. The present derivation is not
sufficiently general to allow for this as it would mean allowing the volumes 
$V_{n}$ to depend on the order parameters. This should be possible but the
results can be expected to be considerably more complex than those presented
here. Such a generalization will be the subject of a future publication.

\begin{acknowledgements}
This work has benefited from conversations I have had with Marc Baus and
Pierre Gaspard. This work was supported in part by the European Space Agency
under contract number C90105.
\end{acknowledgements}

\bigskip 

\end{document}